\documentclass[prl,showpacs,twocolumn]{revtex4}
\usepackage{amsmath}
\usepackage{amsfonts,amssymb,graphicx}

\begin{document}

\title{Closing the gap of secure quantum key rate with the Heralded Pair-Coherent States}
\author{ShengLi Zhang$^{1,2}$, XuBo Zou$^{1}$, ChenHui Jin$^{2}$
and GuangCan Guo$^{1}$}
\affiliation{1 Key Laboratory of Quantum Information, University of Science and
Technology of China (CAS), Hefei 230026, China.\\
2 Zheng Zhou Information and Technology Institute, Zhengzhou, Henan
450004, China}
\date{\today}

\begin{abstract}
In this paper, we investigate the long-standing gap of quantum key
rate between the Weak Coherent Pulse (WCP) and Heralded Single
Photon Sources(HSPS) implementation of quantum cryptographical
protocol. We prove that, by utilizing the Heralded Pair Coherent
State (HPCS) photon sources, such a gap can be actually filled in
both BB84 and SARG quantum key distribution. Thus, a universal
photon source which achieves the up-to-date optimal key rate for
each transmission distance is obtained. 
\end{abstract}

\maketitle

Quantum Key Distribution (QKD)\cite{key1,key3} is a powerful tool
which allows two remote partners, Alice and Bob, to establish random
and private keys with an unconditional security. However, the
unavoidable quantum channel imperfections, such as the high loss of
transmission line and the dark counts of single photon detectors,
impose severe limitations on both key generation rate and
transmission distance of the practical quantum cryptographical
applications.

Recently, to give a solution to these problems, much effort has been
focused on both cryptography protocol itself and experimental
configuration to give a further improvement of the performance of
the state-of-art quantum cryptography system. This includes:
(1)decoy state method where several coherent state with different
intensities are used\cite{Hwang,XBW,HKLOPRL,Harr,XFMa,HKLO2,Cai};
(2)searching for more efficient information reconciliation and
privacy amplification protocol\cite{MoreEDP}; (3)using spatial
freedom of single photon\cite{walborn06} or by utilizing
entanglement photon pairs to transfer more than 1 bit of
information\cite{Silber08,neve05}; (4)exploring more efficient
photon number distribution sources to generate a higher secure key
rate\cite{paris07,lutk,Ou05,PRA2006}, and more experimentally,
(5)applying the high-speed signal modulation and single photon
detection\cite{ghzQKD,HighSpeed}.

Of these efforts, the most efficient and prominent method is to
bring (2) and (4) together, i.e., to use decoy state technology and
explore its application in different kinds of photon source models.
In Ref.\cite{lutk}, N. L\"{u}tkenhaus gave a proposal of the
superiority of the HSPS photon sources in implementing QKD
protocols. Later, following this seminar work, a lot of work has
been done to bring such a topic into a much wider collection of
variations\cite{PRA2006,QinW,ZSL}. It is shown by T. Horikiri
\textit{et al} that the key rate can be improved with a HSPS sources
and the secure distance can be prolonged from 140km to 170km for
BB84 quantum cryptographical protocol\cite{PRA2006}. Enlightened by
such an observation, we gave a further consideration of the
Scarani-Ac\'{\i}n-Ribordy-Gisin(SARG)\cite{SARG} protocol and proved
that such a enhancement from 95km to 120 km for SARG in the
transmission distance can also been obtained \cite{ZSL}.

However, behind all these improvement in QKD practical
implementation, there exist a long-standing problem that has not
been addressed. That is the so-called gap of secure key rate between
WCP and HSPS implementations of quantum cryptographical schemes. As
is shown in Ref.\cite{ZSL}, it is shown that there exists a
threshold distance of 136km for BB84 (93km for SARG) beyond which
the secret key rate is only enhanced and thus, the distance is
greatly prolonged. However, within such a threshold distance,
HSPS+BB84 (HSPS+SARG)has a less secure key generating rate than
WCP+BB84(WCP+SARG). Put simply, to optimize the final performance,
one should resort to WCP sources within the threshold distance but
resort to HSPS sources if the distance is threshold-beyond.

In commercialization of QKD, it is important to find a
\textit{universal} photon-emitting sources which is the optimal in
secure key generating rate for each transmission distance. Then, a
question that naturally arises is that whether there exists a
universal photon sources for QKD applications. In this paper we give
a proof that the answer is actually affirmative.

Our derivation will be given by investigating the performance of
Heralded Pair Coherent State (HPCS) and the corresponding Decoy
State method. In this following, we will first start by introducing
the famous GLLP formula for key generating rate and also the decoy
state method. Then, we will move our attention to the decoy state
implementation of BB84 and SARG protocol with HPCS sources. Finally,
to obtain our conclusion, we give a numerical simulation with the
same experiment parameters taken from \cite{GYS} as in
Ref.\cite{ZSL,PRA2006,HKLO2}.

Different from any classical key distributing protocol, the QKD
protocol, such as BB84 and SARG, bases on its unconditional security
on the fundaments of quantum physics. However, the security proof of
these kind of protocol has not been obtained until many years after
their inventions\cite{sci}. The first simple proof of the single
photon key rate for BB84 is given by Shor and
Preskill\cite{QKDproof}. This proof is then further extended to
explicitly accommodate the imperfections in practical devices, e.g.,
the laser sources which occasionally emits multi-photon signals in
each emitting. It is shown that the secure key rate for imperfect
laser sources can be given by\cite{GLLP}:
\begin{equation}
R_{BB84}\geq -Q_{\mu }f(E_{\mu })H(E_{\mu })+Q_{1}[1-H(e_{1})],
\label{RBB84}
\end{equation}%
where $Q_{\mu }$ is the Gain (the ratio of the number of Bob's
detector counts to the
number of signals Alice emites when they are using the same basis\cite%
{HKLOPRL}) of Bob's detector and $E_{\mu }$ is the bit error rate
(QBER) of
sift key when the average photon number of the laser source is $\mu $. $%
H(x)=-xlog_{2}x-(1-x)log_{2}(1-x)$ is the Binary Shannon function.
The function $f(E_{\mu })$ is the efficiency of key reconciliation
and is often assumed to be an constant 1.22 for
simplicity\cite{HKLO2}. $Q_1,e_1$ is the corresponding gain and
error rate contributed by single-photon proportion of the laser
source.

In the rest of our paper, we will give a thorough analysis of both
BB84 and SARG protocol. So, it is now convenient to introduce the
secure consideration relevant to SARG protocol. For SARG protocol,
it is known that the security key rate $R_{BB84}$
contains the contribution of both single-photon and two-photon pulses\cite%
{HKLO2,SARGproof}
\begin{eqnarray}
R_{SARG}  \nonumber &\geq &-Q_{\mu }f(E_{\mu })H(E_{\mu
})+Q_{1}[1-H(g(e_{1}))]\\&+&Q_{2}\left[ 1-H(h(e_{p,2}))\right]
\label{Sarg}
\end{eqnarray}
in which $e_{p,2}=e_{2}$ is the phase shift error rate contributed
by double-photon pulses and $g(\cdot),h(\cdot)$ are functions that
can be obtained in the Appendix of Ref. \cite{HKLO2}.

In effect, the average gain $Q_\mu$ of laser and the error rate
$E_\mu$ can be directly observed in experiments. To derive the
unconditional secure key rate in Eq. (\ref{RBB84}) and Eq.
(\ref{Sarg}), what one is required to do is only to give an accurate
estimation of $Q_1, e_1, Q_2, e_2$. However, due to strong channel
loss and all kinds of potential and powerful attacks by
eavesdroppers, to obtained the required estimation is not a trivial.
Fortunately, one can apply the decoy state method and continue our
derivation.

The main core of decoy state \cite{Hwang,HKLOPRL,XBW,XFMa} is to
replace the single intensity laser sources with signal states and
decoys states which are generated respectively by lasers with
different intensities. If the signal state and the decoy state have
the same wavelength, timing, and many other physical characters, no
eavesdropper will be able to distinguish a decoy state from a signal
state successfully. Thus the condition probability that Bob's
detector clicks when a $n$-photon pulse is emitted from Alice will
be the same for signal state and decoy state. If we use $Y_{n}$ to
denote such a probability, we have
$Y_{n}(signal)=Y_{n}(decoy)=Y_{n}$ and the error rate of $n$-photon
pulse reads $ e_{n}(signal)=e_{n}(decoy)=e_{n}. $

Typically, if we denote the intensity of signal state by $\mu$, the
decoy state by $\nu_1,\nu_2,\cdots,\nu_k$( Here we assume, $k$
different decoy state are involved.), by averaging over all the
possible $n$-photon pulses, one can obtain that the overall gain
$Q_\mu, Q_{\nu_{k}}$ and over error $EQ_\mu, EQ_{\nu_k}$ follows
\begin{eqnarray}
Q_{\mu}&=&\sum\limits_{n}Y_{n}P(\mu ,n), ~~~~   Q_{\nu_k}
=\sum\limits_{n}Y_{n}P(\nu_k,n), \label{qmu}\\
EQ_{\mu}&=&\sum\limits_{n}e_{n}Y_{n}P(\mu ,n),~ EQ_{\nu_k}
=\sum\limits_{n}e_{n}Y_{n}P(\nu_k,n)\label{eqmu},
\end{eqnarray}
where $P(\mu ,n)$ is the probability that the pulse contains $n$
photon. $E_{\mu }$, $E_{\nu}$ can be obtained by $E_{\mu
}=EQ_{\mu}/Q_{\mu},E_{\nu_k}=EQ_{\nu_k}/Q_{\nu_k}$.


The laser sources that will be utilized in encoding of BB84 or SARG
quantum signal is Heralded Pair Coherent State(HPCS). In fact, the
Pair Coherent State(PCS), which is originally proposed by G. S.
Agarwal\cite{PCS86}, is in essence another kind of photo-number
correlated state. But we will see that its correlation is quite
powerful and could induce a fundamental improvement compared with
the HSPS state. According to Ref.\cite{PCS86,paris07}, the PCS is a
two-mode correlated coherent and can be written in the Fock basis
\begin{eqnarray}
|\mu\rangle\rangle=\frac{1}{\sqrt{I_0(2|\mu|)}}\sum\limits_{n=0}^{\infty}|n\rangle_1|n\rangle_2,
\end{eqnarray}
where $\mu\in\mathbb{C}$ and $I_0(\cdot)$ is the modified Bessel's
function of the first kind. Interestingly, by tracing over arbitrary
one of the two modes, the photon number distribution follows
$\rho_1=\mathrm{Tr}_2[|\mu\rangle\rangle\langle\langle\mu|]=\frac{1}{I_0(2\mu)}\sum_n\frac{\mu^{2n}}{n!}|n\rangle\langle
n|$, which demonstrates a sub-poissonian statistics. To decrease the
potential multi-photon pulses, we consider the photon heralding
technique. After the generation, one mode of the PCS state is sent
to a \textit{trigger detector} (only distinguish click and
non-click), and the other one is sent to the Alice's encoding
module. Only when the trigger detector clicks, will the laser signal
be deemed to have been sent to Bob.
Therefore, when we consider the
the probability that $n$-photon pulse is emitted form Alice enclave,
the quantum detection efficiency $\eta_A$ and also the dark count
rate $d_{A}$ should be included\cite{PRA2006}:
\begin{eqnarray}
P(\mu ,n) &=&\frac{1}{I_0(2\mu)}\frac{\mu
^{2n}}{(n!)^2}\left[1-(1-\eta _{A})^{n}+d_{A}\right].
\label{photonDis}
\end{eqnarray}%

\begin{figure}
  \includegraphics[width=8cm]{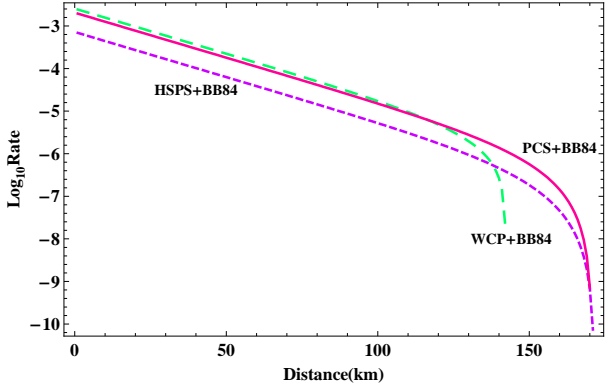}\\
  \caption{Comparison rate of HPCS+BB84 protocol with different photon number distribution, $\eta_A=0.6,d_A=5\times 10^{-8}$. For HSPS+BB84 sources, threshold distance is 136 km.
  The secret rate is optimal
  for each transmission distance.}\label{BB84Rate}
\end{figure}
\begin{figure}
  \includegraphics[width=8cm]{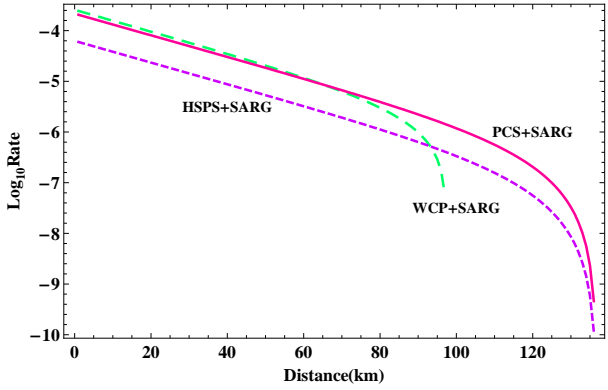}\\
  \caption{Comparison rate of HPCS+SARG protocol with different photon number distribution, $\eta_A=0.6,d_A=5\times 10^{-8}$. For HSPS+SARG sources, threshold distance is 93 km. The secret rate has been optimized
  .}\label{SARGRate}
\end{figure}

With the photon distribution in Eq. (\ref{photonDis}), we will be
able to give a general theory of decoy state method for HPCS-based
quantum key distribution schemes. The decoy state method we present
here is practical in the sense that we can fulfill such a task by
only a finite different number of signal state and decoy state. In
fact, it will be shown that only one signal state $\mu$ and two
decoy state $\nu_1,\nu_2(1>\mu>\nu_1>\nu_2$ and
$\nu_1^2+\nu_2^2\leq\mu^2)$ will be enough to give an estimation.
First of all, it is sufficient to use only $\nu_1,\nu_2$ and obtain
the estimation of $Y_1$ and $e_1$. Plugging Eq. (\ref{photonDis})
into Eq.(\ref{qmu}), one gets
\begin{eqnarray}
Q_{\nu_1} =\sum\limits_{n}Y_{n}\frac{1}{I_0(2\nu_1)}\frac{\nu_1
^{2n}}{(n!)^2}\left[1-(1-\eta _{A})^{n}+d_{A}\right]\label{qv1}\\
Q_{\nu_2} =\sum\limits_{n}Y_{n}\frac{1}{I_0(2\nu_2)}\frac{\nu_2
^{2n}}{(n!)^2}\left[1-(1-\eta _{A})^{n}+d_{A}\right]\label{qv1}
\end{eqnarray}
Form $I_0(2\nu_1)\times Q_{\nu_1}-I_0(2\nu_2)\times Q_{\nu_2}$ , we
have
\begin{eqnarray}
&&I_0(2\nu_1)Q_{\nu_1}-I_0(2\nu_2)Q_{\nu_2}\nonumber\\
&&=Y_1(\eta_A+d_A)+\sum\limits_{n\geq 2}Y_{n}\frac{\nu_1 ^{2n}-\nu_2
^{2n}}{(n!)^2}\left[1-(1-\eta _{A})^{n}+d_{A}\right]\nonumber\\
&&\leq Y_1(\eta_A+d_A)+\frac{\nu_1 ^{4}-\nu_2
^{4}}{\mu^4}\sum\limits_{n\geq
2}Y_{n}\frac{\mu^{2n}}{(n!)^2}\left[1-(1-\eta
_{A})^{n}+d_{A}\right]\nonumber,
\end{eqnarray}
 where in last line we have applied the relation $\nu_1^{2n}-\nu_2^{2n}\leq (\nu_1^4-\nu_2^4)\mu^{2n-4}$ for any $n\geq2$.
 The lower bound for single photon gain $Y_1$ can be obtained:
\begin{eqnarray}
&&Y_1\geq
Y_1^{U,\nu_1,\nu_2}\nonumber\\
&=&\frac{I_0(2\nu_1)Q_{\nu_1}-I_0(2\nu_2)Q_{\nu_2}-\frac{\nu_1^4-\nu_2^4}{\mu^4}\left(I_0(2\mu)Q_{\mu}-d_A
Y_0\right)}{(\eta_A+d_A)\left[(\nu_1^2-\nu_2^2)-\frac{\nu_1^4-\nu_2^4}{\mu^2}\right]}.
\end{eqnarray}
Using a similar way, from $I_0(2\nu_1)\times
EQ_{\nu_1}-I_0(2\nu_2)\times EQ_{\nu_2}$, we have
\begin{eqnarray}
e_1\leq
e_1^{U,\nu_1,\nu_2}=\frac{I_0(2\nu_1)EQ_{\nu_1}-I_0(2\nu_2)EQ_{\nu_2}}{(\eta_A+d_A)Y_1^{L}(\nu_1^2-\nu_2^2)}.
\end{eqnarray}
Furthermore, to derive the two-photon gain $Y_2$ and bit error rate
$e_2$, one need the help of the gain rate for signal state
$Q_{\mu}$:
 \begin{eqnarray} Q_{\mu}
&=&\sum\limits_{n}Y_{n}\frac{1}{I_0(2\mu)}\frac{\mu
^{2n}}{(n!)^2}\left[1-(1-\eta _{A})^{n}+d_{A}\right]\nonumber\\
&=&\sum\limits_{n=0}^{n=2}Y_{n}\frac{1}{I_0(2\mu)}\frac{\mu
^{2n}}{(n!)^2}\left[1-(1-\eta
_{A})^{n}+d_{A}\right]\nonumber\\
&+&\sum\limits_{n=3}^{n=\infty}Y_{n}\frac{1}{I_0(2\mu)}\frac{\mu
^{2n}}{(n!)^2}\left[1-(1-\eta _{A})^{n}+d_{A}\right].\label{qu}
\end{eqnarray}
With some frustrating algebra manipulation, the two-photon gain and
two-photon error rate can be given by
\begin{widetext}
\begin{eqnarray}
Y_{2} &\geq &{Y_{2}}^{Est}\nonumber \\
&=&\frac{2}{[1-(1-\eta
_{A})^{2}+d_A][(a^{2}-b^{2})-\frac{a^{3}-b^{3}}{c}]}
[I_0(2\nu_1)Q_{\nu_{1}}-I_0(2\nu_{2})Q_{\nu_{2}}-\frac{a^{3}-b^{3}}{c^{3}}I_0(2\mu)Q_{\mu}+\nonumber \\
&&+Y_{0}d_{A}\frac{a^{3}-b^{3}}{c^{3}}-Y_{1}\eta _{A}[(a-b)-\frac{a^{3}-b^{3}%
}{c^{2}}]].\\
e_{2}&<&{e_{2}}^{Est}=\frac{\nu_2^2 I_0(2\nu_1)EQ_{\nu_{1}}-\nu_1^2I_0(2\nu_2)EQ_{\nu_{2}}+\frac{1}{2}%
(\nu_1^2-\nu_2^2)Y_{0}d_{A}}{Y_{2}[1-(1-\eta
_{A})^{2}+d_A]\nu_1^2\nu_2^2(\nu_1^2-\nu_2^2)}.
\end{eqnarray}
\end{widetext}

Now let's give some numerical simulation with the experimental
parameters taken from Ref.\cite{GYS}. To move on, we use the
$Q_{\mu},EQ_{\mu}$ in the ideal scenario to replace the actual
experiment and investigate the limitations of the key generation
rate with HPCS. In precise, for a $n$-photon pulse which
is not disturbed by eavesdropper, the expectation values of yields
and bit error rate can be respectively evaluated\cite{HKLO2}. For
BB84:
\begin{eqnarray}
Y_{n,BB84} &=&\left[\eta _{n}+(1-\eta_{n})p_{dark}\right]/2,  \label{ynbb84} \\
e_{n,BB84}&=&\frac{\eta _{n}\frac{e_{\text{det}}}{2}%
+(1-\eta _{n})p_{\text{dark}}\frac{1}{4}}{Y_{n,BB84}},
\label{eynbb84}
\end{eqnarray}%
and for SARG:
\begin{eqnarray}
Y_{n,SARG} &=&\eta _{n}\left(
\frac{e_{\text{det}}}{2}+\frac{1}{4}\right)
+\left( 1-\eta _{n}\right) p_{\text{dark}}\frac{1}{2},  \label{ynsarg} \\
e_{n,SARG}&=&\frac{\eta _{n}\frac{e_{\text{det}}}{2}%
+(1-\eta _{n})p_{\text{dark}}\frac{1}{4}}{Y_{n,SARG}},
\label{ensarg}
\end{eqnarray}%
where $p_{dark}$ is the dark count Bob's detector and $e_{\text{det
}}$stands for the misalignment of the optical instrument.
$\eta_n=1-(1-\eta)^n$ and $\eta $ denotes the total transmittance of
quantum channel and Bob's enclaves.

Substituting Eqs. (\ref{ynbb84})(\ref{eynbb84}) and
Eq.(\ref{photonDis}) into Eqs. (\ref{qmu})(\ref{eqmu}), we get
$Q_{\mu,BB84}=\frac{1}{2}\xi-(1-p_{dark})\zeta,
EQ_{\mu,BB84}=\frac{e_{det}}{2}\xi-\left(\frac{e_{det}}{2}-\frac{p_{dark}}{4}\right)\zeta,
E_{\mu,BB84}=EQ_{\mu,BB84}/Q_{\mu,BB84}. $ Similarly, for SARG
protocol, we obtain
$Q_{\mu,SARG}=\left(\frac{e_{det}}{2}+\frac{1}{4}\right)\xi-\left(\frac{e_{det}}{2}+\frac{1}{4}-\frac{p_{dark}}{2}\right)\zeta,
EQ_{\mu,SARG}=\frac{e_{det}}{2}\xi-\left(\frac{e_{det}}{2}-\frac{p_{dark}}{4}\right)\zeta,
E_{\mu,SARG}=EQ_{\mu,SARG}/Q_{\mu,SARG}^,$ where $\xi
=1+d_A-\frac{I_0(2\mu\sqrt{1-\eta_A})}{I_0(2\mu)}, \zeta
=(1+d_A)\frac{I_0(2\mu\sqrt{1-\eta})}{I_0(2\mu)}-\frac{I_0(2\mu\sqrt{(1-\eta)(1-\eta_A)})}{I_0(2\mu)}
$.

This ideal scenario is a good indication of the limitation of the
performance of secure key rate. To see this, the single error $e_1$
and $Q_1$ is obtained directly from Eq.(\ref{eynbb84}) and
Eq.(\ref{ensarg}). The parameters for trigger detector are
$\eta_A=0.6,d_A=5\times 10^{-8}$ as the ones chosen in Ref.
\cite{PRA2006}. The average photon number for signal state $\mu$ is
optimized for each transmittance distance. As is shown in Fig
.\ref{BB84Rate}, the original threshold distance for BB84 at 136km,
doesn't exist for HPCS state and an average gain of $1.5$ in key
generating rate can be observed. Fig. \ref{SARGRate}, the SARG
protocol within different implementation of photon sources is
depicted. Comparison with the BB84, the average gain is much more
prominent, which is about $3.4$ times than ever before.

 In summary, by investigating the sub-possion
photon sources, we show the gap of secrete key rate between the WCP
and HSPS implementations can be actually filled. For the Heralded
Pair Coherent State, the threshold distance does not exist and can
be considered as a convenient and universal photon surces in QKD
commercializations. However, we note that here, to look for a
convenient and efficient photon sources is still an open an
important question for the future robust and high speed QKD
application.

After finishing our task we are aware that, more Recently, the
generation of pair coherent state has also been addressed in
Ref\cite{genePCS}.

S. Z. thanks Prof. QingYu Cai and  Dr. XuanYang Lai from the Wuhan
Institute of Physics and Mathematics(CAS) for their warm and kind
accommodation during his summer visit. This work was supported by
National Fundamental Research Program, also by National Natural
Science Foundation of China (Grant No. 10674128 and 60121503) and
the Innovation Funds and \textquotedblleft Hundreds of
Talents\textquotedblright\ program of Chinese Academy of Sciences
and Doctor Foundation of Education Ministry of China (Grant No.
20060358043).

\end{document}